\documentclass[traditabstract, a4paper]{aa}

\usepackage{amsmath}
\usepackage{txfonts}
\usepackage[T1]{fontenc}
\usepackage[latin9]{inputenc}
\usepackage{amstext}
\usepackage{graphicx}
\usepackage{esint}
\usepackage[authoryear]{natbib}
\usepackage{float}
\usepackage{multirow} 
\usepackage{rotating}



\begin{document}

\title{Improved CLEAN reconstructions for rotation measure synthesis with maximum likelihood estimation}

\author{M.R.~Bell\thanks{\email{mrbell@mpa-garching.mpg.de}}\inst{\ref{inst:mpa}} 
	\and N.~Oppermann\inst{\ref{inst:mpa}} 
	\and A.~Crai\inst{\ref{inst:jacobs}} 
	\and T.A.~En\ss lin\inst{\ref{inst:mpa}}}

\institute{Max Planck Institute for Astrophysics, Karl-Schwarzschild-Str. 1, 85748 Garching, Germany\label{inst:mpa} 
	\and Jacobs University, Campus Ring 1, 28759 Bremen, Germany\label{inst:jacobs}}

\date{Received \textbf{???} / Accepted \textbf{???}} 

\abstract{The CLEAN deconvolution algorithm has well-known limitations due to the restriction of locating point source model components on a discretized grid. In this letter, we demonstrate that these limitations are even more pronounced when applying CLEAN in the case of Rotation Measure (RM) synthesis imaging. We suggest a modification that uses Maximum Likelihood estimation to adjust the CLEAN-derived sky model. We demonstrate through the use of mock one-dimensional RM synthesis observations that this technique shows significant improvement over standard CLEAN and gives results that are independent of the chosen image pixelization. We suggest using this simple modification to CLEAN in upcoming polarization sensitive sky surveys.}


\titlerunning{Improved RMCLEAN reconstructions}
\authorrunning{M.R.~Bell et al.}

\maketitle

\section{Introduction\label{sec:Introduction}}

CLEAN \citep{1974_hogbom} is a deconvolution algorithm that is widely used throughout astronomy. In CLEAN, a deconvolved image is constructed iteratively by locating the maximum in the image, adding a delta function at that location with some fraction of the peak brightness to a model, and then subtracting a point spread function (PSF) that has been shifted to the location of the model component and scaled by its strength. This is repeated until a user-defined stop condition has been reached. 

The algorithm makes an implicit assumption that the image is well described by a collection of statistically independent point sources. In some cases this is an adequate assumption and experience has shown that even in circumstances where this assumption is not ideal, CLEAN is still able to produce qualitatively reasonable results. Because of its flexibility, simplicity, and speed, CLEAN has become the standard tool for image reconstruction in radio astronomical aperture synthesis imaging.

A limitation of CLEAN arises due to the discretization of the image space, and therefore one can only create model point sources on a regularly spaced grid. To properly remove the PSF pattern associated with a strong point source, it is crucial for the delta function representing the source in the CLEAN model to have an accurate position. While the PSF limits the minimum spatial scale of structures that can be resolved, the accuracy with which one can measure the location of a point source depends on the signal-to-noise ratio. In the high signal to noise regime the uncertainty in the location of a source can be much smaller than the PSF, and to measure the source location accurately with a single model point source one would have to create an image with a large number of pixels within the PSF. Such over-resolving is at best computationally wasteful and in some cases may make the imaging problem completely unfeasible. For example, even with a few pixels per resolution element, three-dimensional (3D) Faraday synthesis \citep{bell_2012} images will easily exceed the available memory on a single computer.

CLEAN is able to partially overcome this issue by describing a single point source by a cluster of CLEAN components around the true source location. The strongest
model point source will lie on the pixel closest to the source location.
Weaker model points will be added to neighboring grid points such that when the collection of model sources is convolved by a Gaussian restoring beam that approximates the idealized
PSF, the resulting image will be closer to the true source position and flux than that of a single model point located on the grid. Nevertheless, inaccuracies in image reconstructions due to the pixelization of the sky are a well known limiting factor to dynamic range when using CLEAN \citep{briggs_1992, perley_1999, cotton_pixelization-in-interferometry_2008, yatawatta_fundamental_2010}.

Rotation Measure (RM) synthesis \citep{brentjens_faraday_2005} is a promising new technique for studying magnetic fields with the new generation of broadband radio telescopes. RM synthesis allows for the separation of multiple sources of polarized emission along a line of sight when each is Faraday rotated by different amounts. It produces an estimate of the Faraday spectrum, or the polarized emission as a function of the Faraday depth, which is a quantity that measures the amount of Faraday rotation. The Faraday depth, $\phi$, is proportional to $\int n_\mathrm{e} B_z \mathrm{d}z$ where $n_\mathrm{e}$ is the density of thermal electrons and $B_z$ is the component of the magnetic field along the line of sight. RM synthesis is similar to aperture synthesis imaging due to the Fourier relationship between the Faraday spectrum and the polarized intensity as a function of the squared wavelength. CLEAN has been proposed as a deconvolution technique to be applied to RM synthesis imaging \citep{heald_westerbork_2009}, and is often referred to as RMCLEAN in this context. Several alternative image reconstruction techniques have been proposed \citep{frick_wavelet-based_2010, li_compressed-sensing-rmsynth_2011, andrecut_sparse-rmsynth_2011}, but RMCLEAN has thus far been used most often as the deconvolution method of choice \citep{feain_2009, harvey-smith_2010, mao_2010, bernardi_2010, brentjens_2011, vaneck_2011, heald_2012, mao_2012b, mao_2012, vanweeren_2012, iacobelli_2012}.

In this letter we show that the limitations of CLEAN due to pixelization of the sky are particularly pronounced in the case of RM synthesis, and propose to use a Maximum Likelihood (ML) estimation procedure to improve the model obtained using the standard RMCLEAN algorithm. This method is similar to those proposed previously in the context of high-fidelity aperture synthesis imaging \citep{el-behery_1980, yatawatta_fundamental_2010, bernardi_subtraction-of-point-sources-in-radio-astro_2011}. It is also similar to the method proposed by \citet{osullivan_2012}, but is more easily scaled to higher dimensions and includes fewer assumptions about the Faraday spectrum. Our aim is to maintain or improve accuracy of the standard RMCLEAN algorithm while keeping the number of pixels to a minimum. 

\section{The algorithm\label{sec:The-algorithm}}

Here we consider the case of 1D RM synthesis where one is attempting to reconstruct the complex valued Faraday spectrum, $s$, from some polarized sky brightness data, $d$, that has been taken at many frequencies. We note that the following description is trivially extended to 3D as would be required in the case of Faraday synthesis \citep{bell_2012}, which combines aperture and RM synthesis into a single procedure that provides significantly better results than when they are performed separately. 

The Faraday depth coordinate axis (the image coordinate) will be denoted as $\phi$, and the data coordinate is $\lambda^2$. For RM synthesis, the following measurement equation applies:
\begin{equation}
d(\lambda^2)=S(\lambda^2)\int \mathrm{d}\phi s(\phi) \mathrm{e}^{2i\phi\lambda^2}+n(\lambda^2).\label{eq:measurement_eq}
\end{equation}
Here $n$ is a Gaussian additive noise term and $S$ represents a sampling function defining the discrete $\lambda^2$ values at which measurements are obtained.

Using the CLEAN algorithm, one generates a list of model points $M={m_i, \phi_i}_i$ with
flux values and positions given by $m_{i}$ and $\phi_{i}$, respectively,
where $i$ is an index over the list of $N_{M}$ model points. The
representation of the model in data space is
\begin{equation}
\widetilde{d_{j}}=\sum_{i}m_{i}e^{2 i\phi_{i}\lambda^2_{j}},\label{eq:model_in_data-space}
\end{equation}
where $j$ is an index over the $N_{d}$ values of $\lambda^2$ for which measurements have been made.

We assume that the probability of measuring the full data set, $d$,
given the CLEAN model (i.e. the likelihood), is a product of the probabilities
of measuring each individual data point, $d_{i}$. In this way, we
assume the measurements to be independent of one another, thus making the
likelihood 
\begin{equation}
P(d|M)=\prod_{j}P(d_{j}|M)=\prod_{j}\sqrt{\frac{1}{2\pi \sigma_j^2}}e^{-\frac{1}{2\sigma_j^2}\left(d_{j}-\widetilde{d_{j}}\right)^2}.
\end{equation}
Here, as mentioned previously, we assume Gaussian distributed noise, and $\sigma_{j}^2$ is the variance of the noise in channel $j$. 


Our approach is to adjust the CLEAN model by maximizing the likelihood
with respect to $m_{i}$ and $\phi_{i}$. For simplicity,
we opt to work with the negative log-likelihood 
\begin{equation}
H(d|M)=-\textrm{ln}\, P(d|M)=\frac{1}{2}\sum_{j}\left[\frac{\left(d_{j}-\widetilde{d_{j}}\right)^{2}}{\sigma_{j}^{2}}+\textrm{ln}\left(2\pi\sigma_{j}^{2}\right)\right].\label{eq:log_likelihood}
\end{equation}
The maximum likelihood solution is obtained by \textit{minimizing} this function. There are many possible approaches to minimizing
Eq.~\ref{eq:log_likelihood}. The iterative procedure that we use is presented in Appendix~\ref{sec:Maximization-approach}.

Minimization of Eq.~\ref{eq:log_likelihood} provides an estimate for the optimal model parameters under the assumption that the number of point sources in the initial CLEAN model is appropriate. This, however, is almost certainly not the case. CLEAN will naturally over-estimate the number of point sources that are required to model the data because it needs a cluster of sources in order to model a single source at an arbitrary location. 

To find the most appropriate number of free parameters in the model that are supported by the data, we modify $H$ to add a term that penalizes additional degrees of freedom. We use the so-called Bayesian Information Criterion \citep[BIC]{schwarz_1978} to suit this role. The BIC, $C$, is given by

\begin{equation}
	C = 2 H(d|m) + 3 N_{m} \mathrm{ln} N_{d}.
\label{eq:BIC}
\end{equation}

The algorithm proceeds as follows: 
\begin{enumerate}
\item Start with a list of model point sources generated by RMCLEAN. Condense the model such that there is only one entry per pixel location.
\item Throw away any model entries with a flux below a user-defined threshold. This step removes model points that result from cleaning too deeply or that only function to slightly shift a single point source location. We use a threshold of twice the CLEAN image noise level. This step is not necessary, but can speed up computation in case there are many excess point sources in the initial model.
\item Take one or more steps to iteratively minimize Eq.~\ref{eq:log_likelihood} for the current model using e.g. the prescription outlined in Appendix~\ref{sec:Maximization-approach}.\label{it:min_H}
\item Attempt to merge nearby model points. Pairwise combine CLEAN components into a single component having a location at the flux-weighted mean location and a flux that is either the sum of the two model point fluxes or that is solved for using e.g. Eq.~\ref{eq:delta_m_solutions}. Accept the merged CLEAN component if $C$ is reduced, otherwise revert to the previous model.\label{it:test_merger}
\item Iterate steps \ref{it:min_H} and \ref{it:test_merger} until the fractional change in $H$ from one step to the next is below a user-defined threshold.
\item Obtain a residual image by subtracting the new model from the data.
\item Convolve the new model with an idealized PSF and add it to the residual image.
\end{enumerate}

\begin{figure}[t]
	\begin{centering}
	\resizebox{\hsize}{!}{\includegraphics{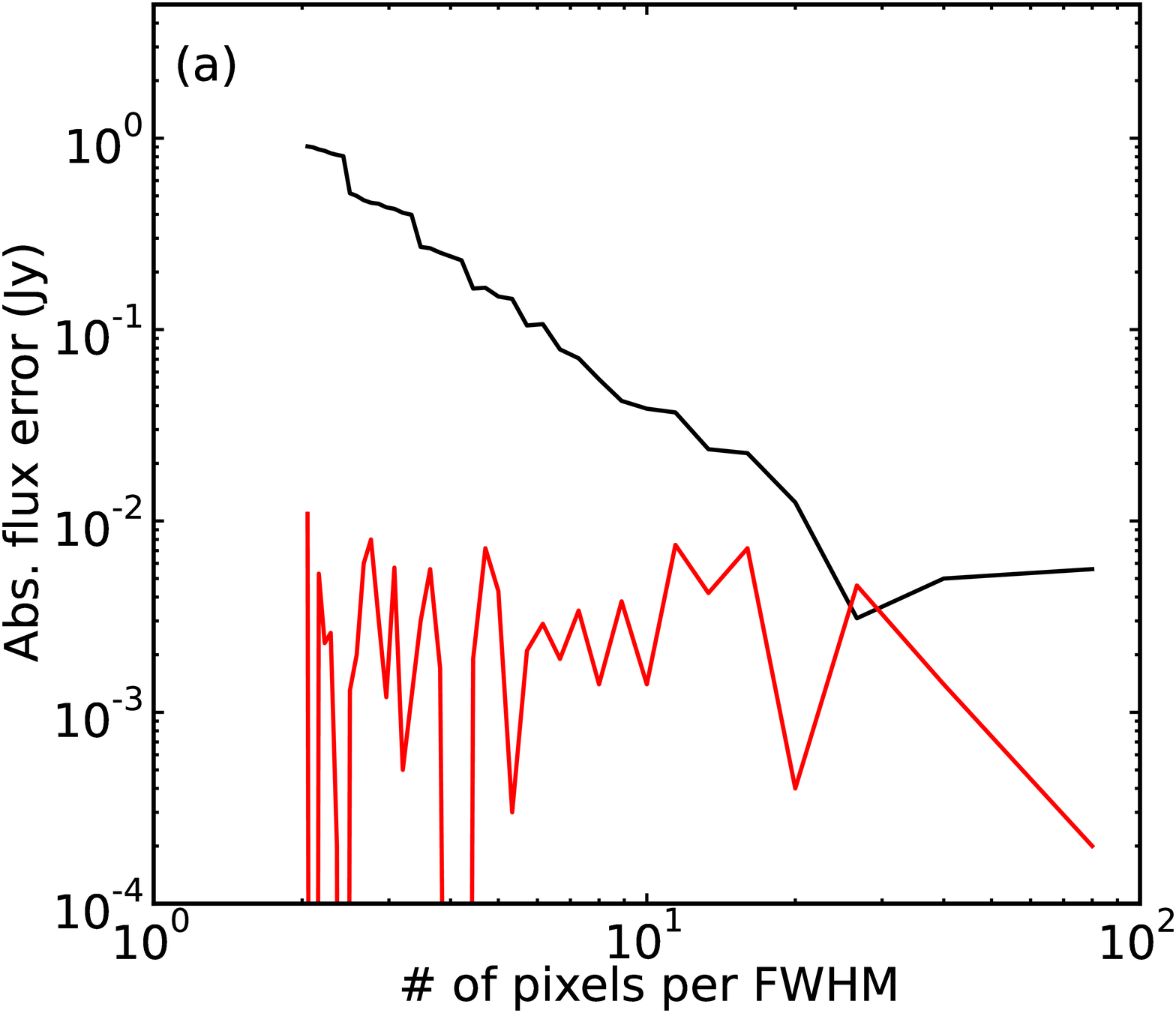} \includegraphics{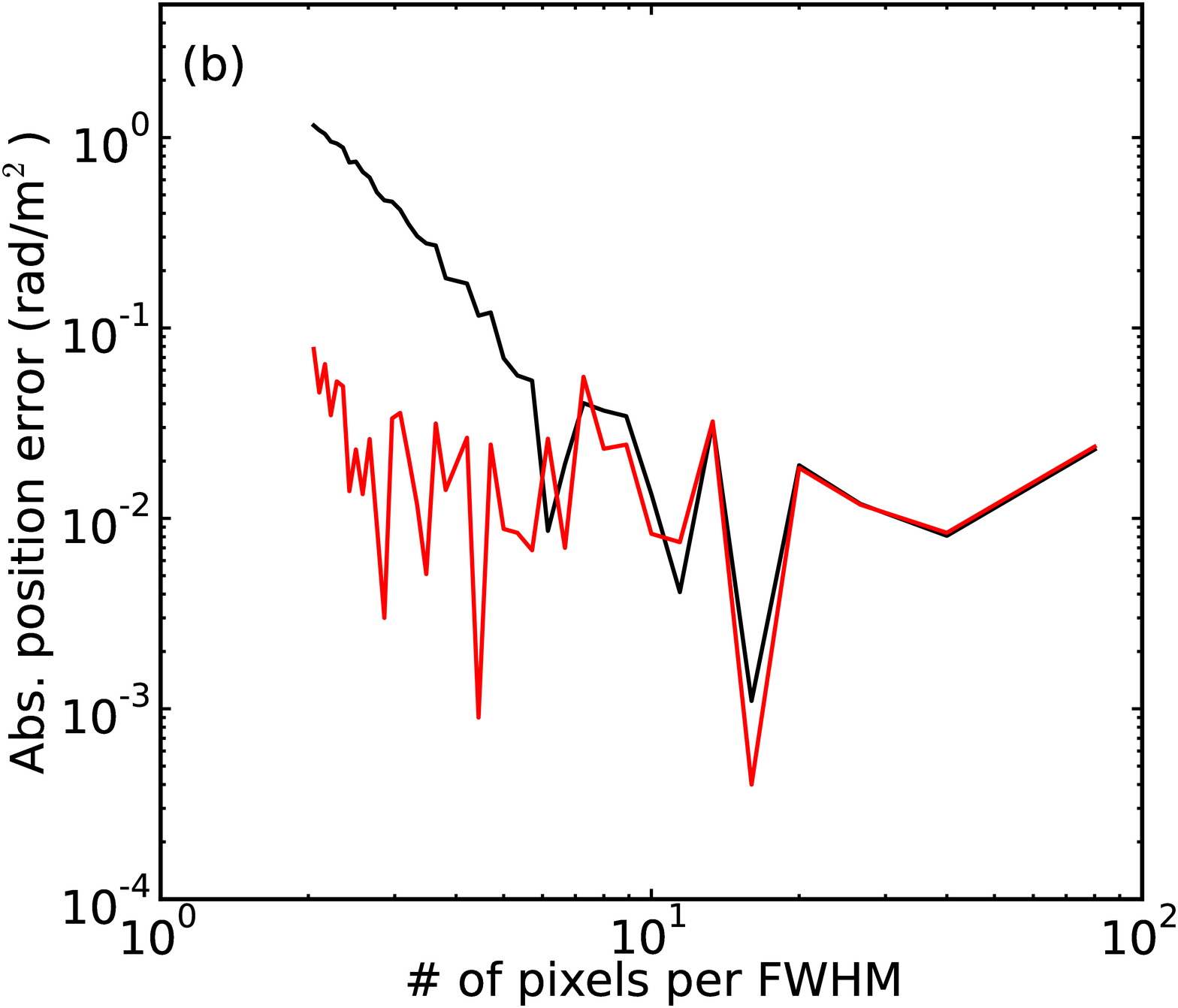}}
	\par\end{centering}
	\caption{A comparison of the (a) flux and (b) position errors as a function of image pixel size in reconstructions of a single point source using RMCLEAN (black line) and the ML algorithm described in this paper (red line). The reported errors are the average absolute errors from 100 simulated data sets. \label{fig:singlesouce_varying_res} }
\end{figure}


\begin{figure}[t]
	\begin{centering}
	\resizebox{\hsize}{!}{\includegraphics{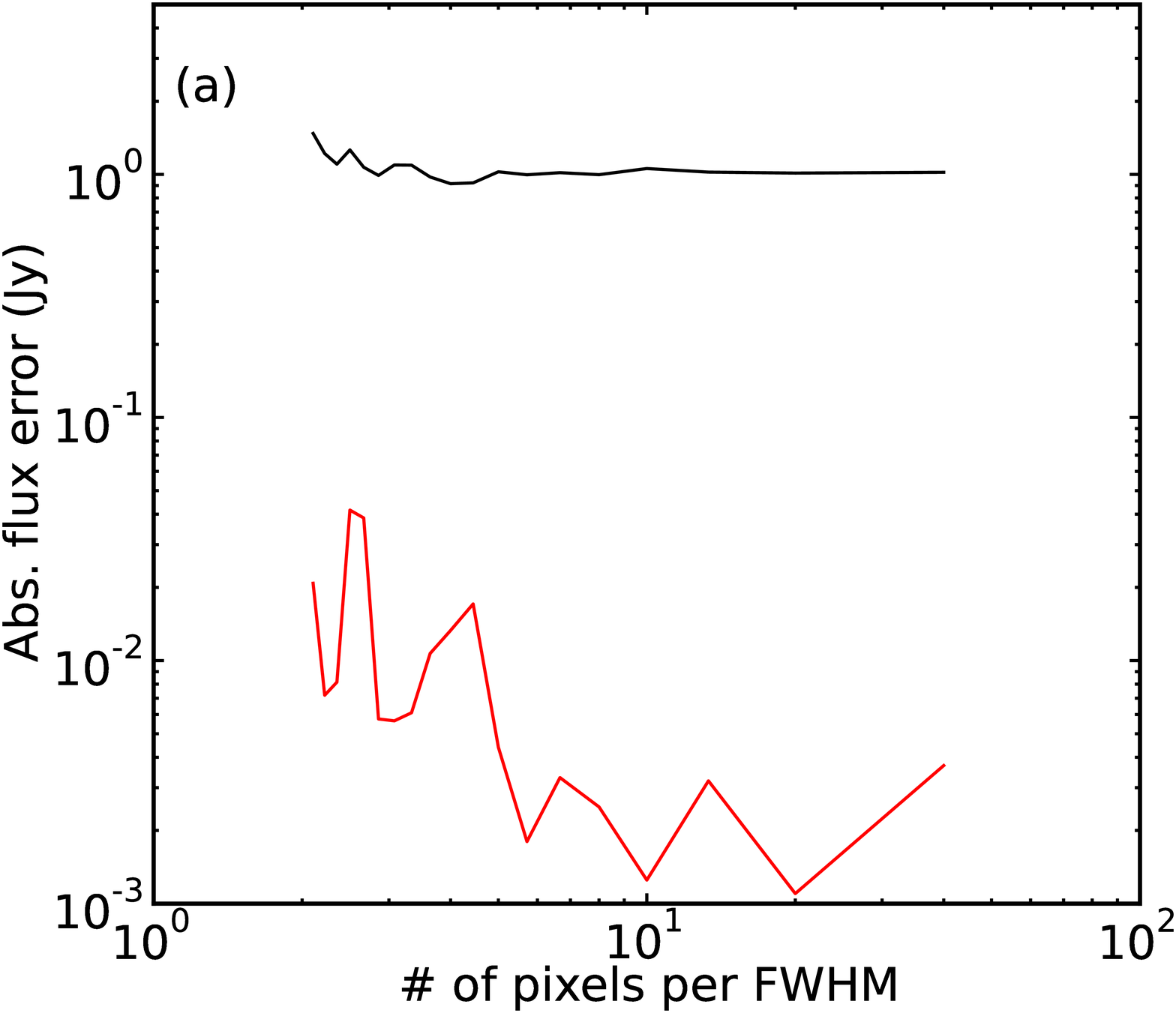} \includegraphics{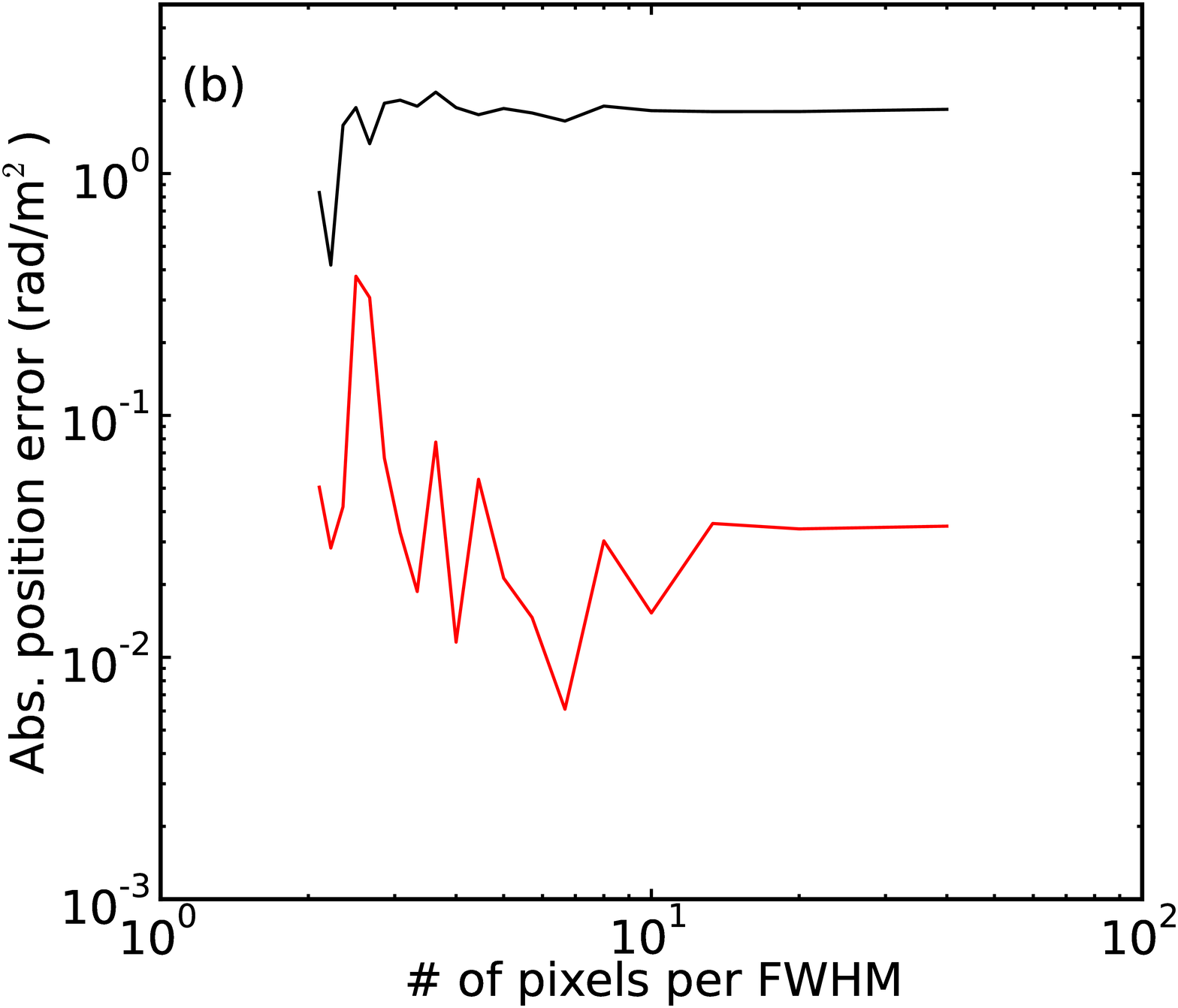}}
		\resizebox{\hsize}{!}{\includegraphics{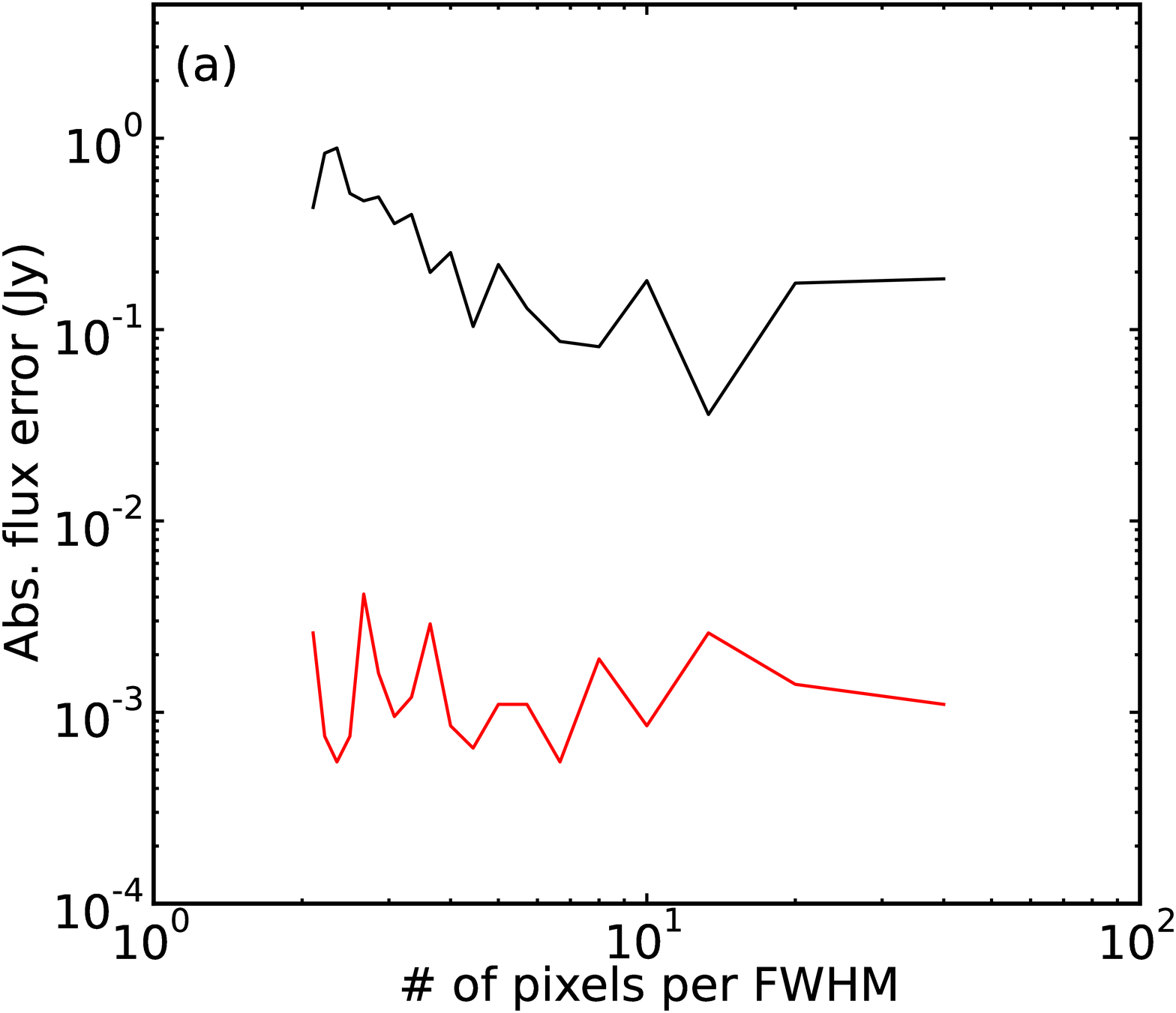} \includegraphics{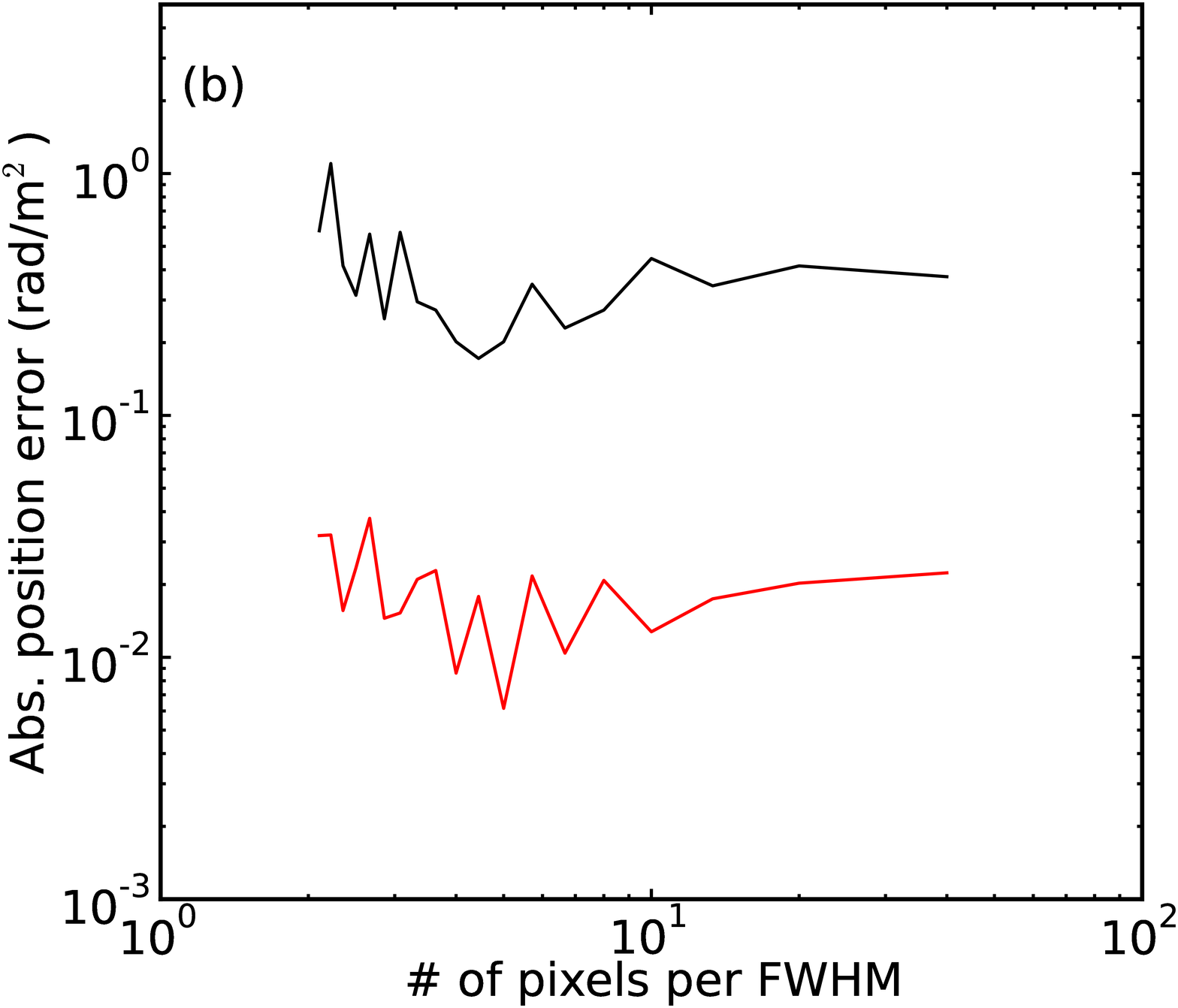}}
	\par\end{centering}
	\caption{As Fig.~\ref{fig:singlesouce_varying_res} but with the average errors of two sources that are separated by 2$\times$FWHM (top row) and 6$\times$FWHM (bottom row). \label{fig:twosouce_varying_res} }
\end{figure}

\section{Demonstration\label{sec:demonstration}}

To demonstrate the effectiveness of the approach outlined above, we present a series of 1D RM synthesis mock observations of simple single and double point-source sky models. We generate simulated data according to Eq.~\ref{eq:measurement_eq} including Gaussian random white noise and sample using a frequency coverage with 800 channels equally spaced between 1 and 4.2~GHz. This frequency range is similar to that of the combination of the upgraded Very Large Array L- and S-band receivers and corresponds to a PSF with a full width at half maximum (FWHM) of the main peak of $\sim40$~rad/m$^2$. A Clark style CLEAN \citep{1980_clark} is then performed and the resulting model is used as input for the ML algorithm. The output images from both procedures are stored for comparison.

In Fig.~\ref{fig:singlesouce_varying_res} we compare the performance of standard RMCLEAN to that of the ML method at reconstructing the position and flux of a single point source as a function of pixel size. The source is located at roughly the same $\phi$ location at each resolution, but shifted slightly to ensure that it is always located one-third of the way between two pixels. The source flux is roughly 100 times the noise level. The plotted flux and position errors are the average of the magnitude of those from 100 trials with different noise realizations. Source position and flux are measured by locating the maximum of the modulus of the Faraday spectrum and fitting a Gaussian to the image around this point. We find that for RMCLEAN the position and flux errors increase significantly with pixel size while those for the ML approach remain nearly constant. Six or more pixels per FWHM of the PSF are required before the positional accuracy for RMCLEAN matches that of the ML approach. To obtain accurate fluxes with RMCLEAN, 40 or more pixels per FWHM are required.

For a sky model with a single source, RMCLEAN is found to work well as long as the image is sufficiently over-resolved. With more than one source, however, over-resolution is no longer effective, as shown in Fig.~\ref{fig:twosouce_varying_res}. Here we show positional and flux errors in the case where there are two sources along the line of sight. In one case, they are separated by 80~rad/m$^2$ (roughly 2$\times$FWHM of the PSF) and in the other by 240~rad/m$^2$ (6$\times$FWHM). The modulus and phase of the complex valued flux is the same for both sources. The errors that are plotted are the average of those from both sources (the errors for the individual sources were qualitatively similar). We find that increasing the resolution does not reduce errors in the RMCLEAN reconstruction and that on average these errors are two or more orders of magnitude larger than in the maximum likelihood reconstructions. This behavior is due to the interactions between the highly structured, complex valued PSFs associated with each point source. The exact dependence on the error as a function of source separation depends on the specific sampling function, and therefore PSF, being used.

\begin{figure}[t]
	\begin{centering}
	\resizebox{\hsize}{!}{\includegraphics{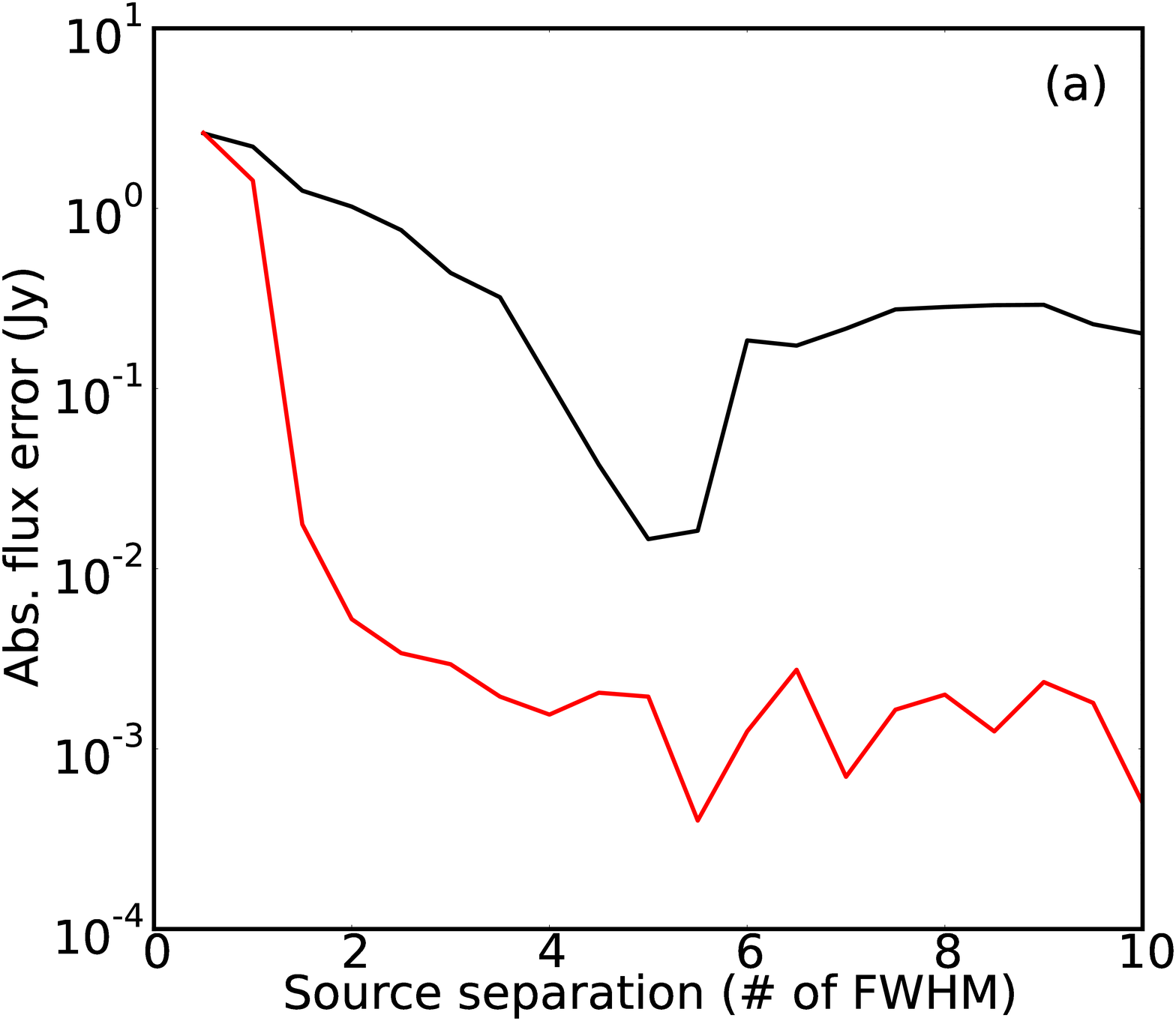} \includegraphics{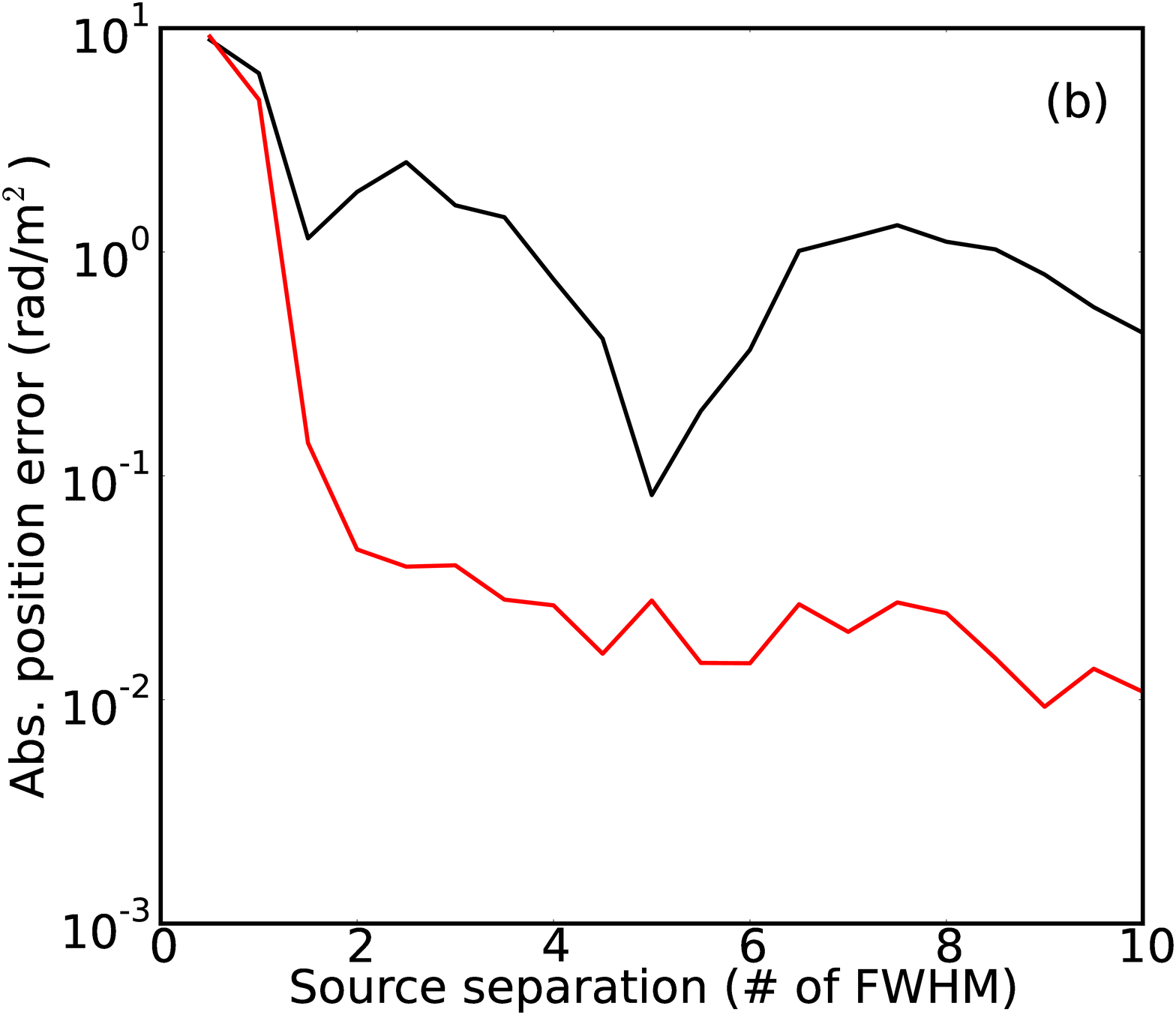}}
	\par\end{centering}
	\caption{As Fig.~\ref{fig:twosouce_varying_res} but with fixed resolution and varying source separation, which is given as a multiple of the FWHM of the PSF (40 rad/m$^2$). \label{fig:varying_sep} }
\end{figure}

We also find that the errors are larger for both approaches when sources are close together relative to the FWHM of the PSF. To test this further, we kept the image resolution fixed (at 5~rad/m$^2$ per pixel) and varied the separation between the two sources (again, each having the same flux and phase). In Fig.~\ref{fig:varying_sep} we see that below a source separation of 2$\times$FWHM the errors for both methods increase dramatically. Nevertheless, at separations of 1$\times$FWHM or larger, the maximum likelihood approach fares significantly better than RMCLEAN. 

\begin{figure*}[t]
	\begin{centering}
	\resizebox{0.9\hsize}{!}{\includegraphics{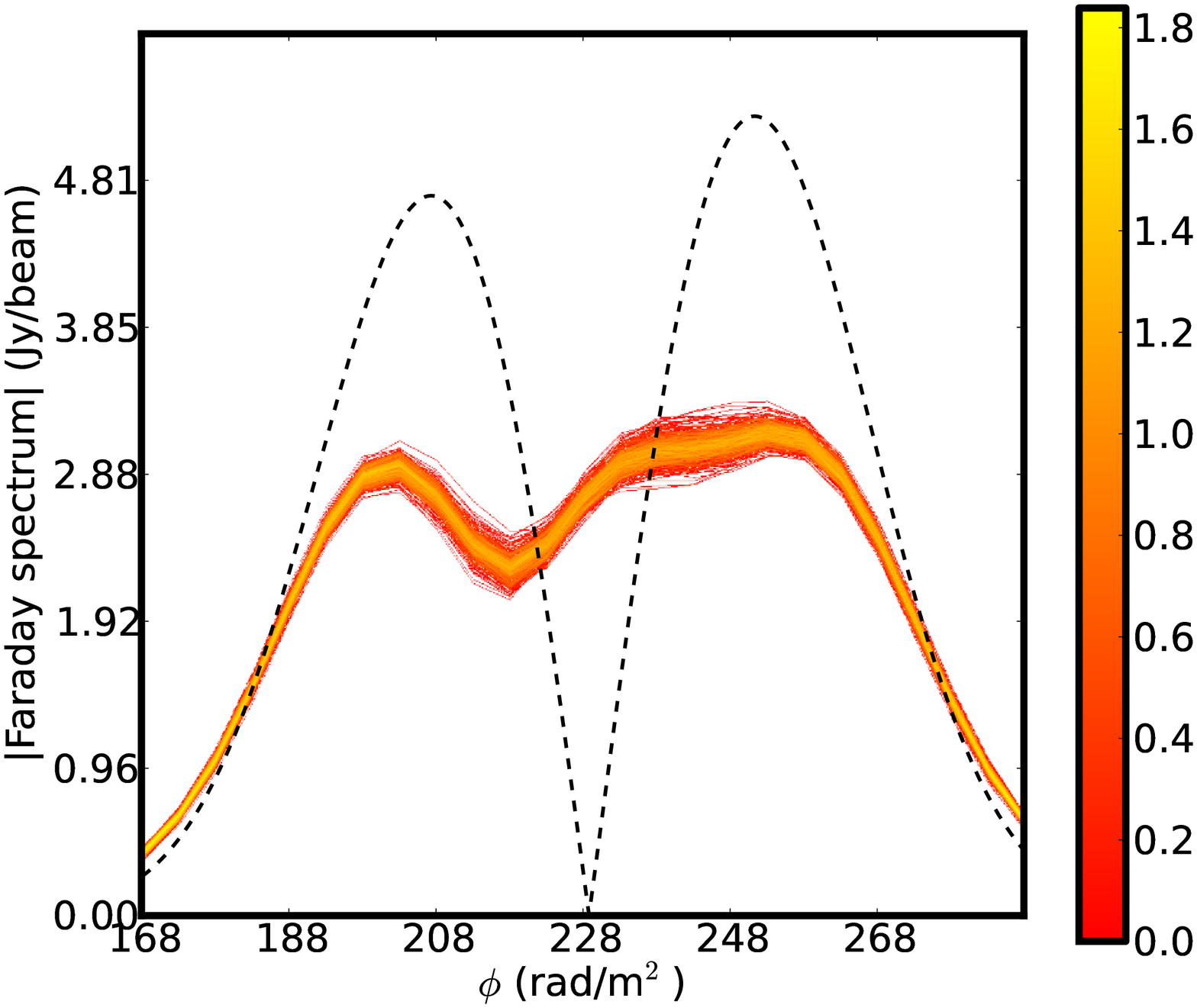}\hspace{3cm} \includegraphics{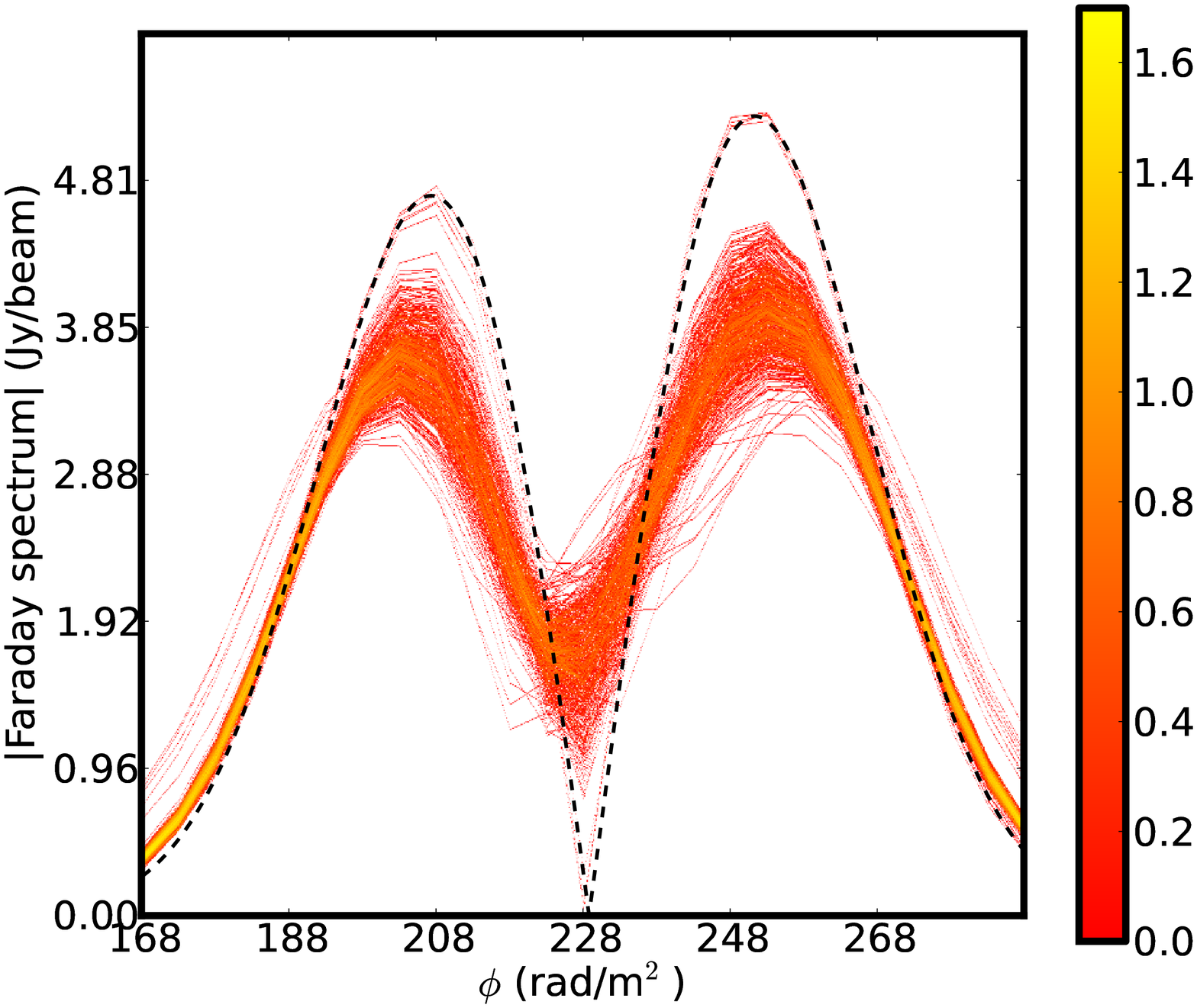}}
	\par\end{centering}
	\caption{Modulus of the Faraday spectra of two point sources having similar fluxes and the same phase and that are separated by 1$\times$FWHM (40~rad/m$^2$). The dashed line shows the sky model after convolution with the idealized PSF. The (a) RMCLEAN and (b) ML reconstructions from 500 trials are plotted in red/yellow, with the color scale indicating the log of the number of reconstructions that pass through each location on the figure.\label{fig:plot_density_plot} }
\end{figure*}

Even though the errors increase dramatically as source separation decreases for both methods, we find that the ML reconstructions are still a significant improvement over RMCLEAN. In Fig.~\ref{fig:plot_density_plot} we show reconstructed Faraday spectra from 500 data realizations of the same sky model. Although the ML results show a systematic reduction in flux and a shift in position relative to the sky model, two distinct sources are clearly identifiable. This is not the case in the RMCLEAN image. We performed similar tests for hundreds of combinations of source separation (between 1 and 6$\times$FWHM), relative fluxes (flux ratios between 1 and 10), relative phases, and noise levels. In all cases the ML approach was an improvement over RMCLEAN, and it only showed significant errors in a few cases like the one shown in Fig.~\ref{fig:plot_density_plot}.

\section{Conclusions}\label{sec:conclusions}

We have shown that the well-known limitations of the CLEAN algorithm that arise due to discretization of the sky are particularly pronounced in the case of RM synthesis. We find that accuracy of the results obtained by CLEAN depend strongly on the choice of pixelization, and significant over-resolution is required to obtain accurate measurements of flux and Faraday depth in the simplest case where there is a single source along the line of sight. Furthermore, RMCLEAN is unable to accurately reconstruct fluxes and locations of the individual sources even with over-resolution when there are multiple sources of emission. 

We propose a simple algorithm to adjust the CLEAN model parameters using Maximum Likelihood estimation together with a prescription for reducing the degrees of freedom in the model to the minimum number that are supported by the data. We show that this algorithm improves upon the results obtained using RMCLEAN alone and provides highly accurate reconstructions independent of the choice of pixelization of Faraday depth space. 

Both algorithms struggle in cases where two point sources are located within 1$\times$FWHM of the PSF, but this merely reflects the fundamental limitation of the observations to resolve structures on these scales. However, although errors increase as source separation decreases, the ML algorithm still provides significant improvement over RMCLEAN. 

CLEAN is ideally suited for the case where the image to be reconstructed is well-described by a set of independent point sources, but as we have shown, does not perform well even in these ideal circumstances when applied in the case of RM synthesis. The algorithm described herein provides significant improvements over RMCLEAN alone and is easy to append to existing RM synthesis imaging pipelines. Therefore, we recommend its use in upcoming polarization surveys that plan to include RM synthesis imaging.

The ML procedure still assumes that the image is described by a set of independent point sources, and will likely not be ideal in case large-scale, diffuse emission is present. A point-source based image reconstruction method is nevertheless an important tool to have on hand. It can be used as a method for separating point sources from a diffuse background, which may be better reconstructed using other methods that do not handle point sources well. It may also be the optimal deconvolution method for use with instruments that are not sensitive to scales much larger than the PSF, as is the case with e.g. LOFAR \citep{beck_2012}.

\begin{acknowledgements}

This research was performed in the framework of the DFG Forschergruppe 1254 Magnetisation of Interstellar and Intergalactic Media: The Prospects of Low-Frequency Radio Observations. We thank Henrik Junklewitz and Marco Selig for many helpful discussions.

\end{acknowledgements}

\bibliographystyle{aa}
\bibliography{rmsynth_refs}

\begin{thebibliography}{28}
\expandafter\ifx\csname natexlab\endcsname\relax\def\natexlab#1{#1}\fi

\bibitem[{{Andrecut} {et~al.}(2012){Andrecut}, {Stil}, \&
  {Taylor}}]{andrecut_sparse-rmsynth_2011}
{Andrecut}, M., {Stil}, J.~M., \& {Taylor}, A.~R. 2012, \aj, 143, 33

\bibitem[{{Beck} {et~al.}(2012){Beck}, {Frick}, {Stepanov}, \&
  {Sokoloff}}]{beck_2012}
{Beck}, R., {Frick}, P., {Stepanov}, R., \& {Sokoloff}, D. 2012, \aap, 543,
  A113

\bibitem[{{Bell} \& {En{\ss}lin}(2012)}]{bell_2012}
{Bell}, M.~R. \& {En{\ss}lin}, T.~A. 2012, \aap, 540, A80

\bibitem[{{Bernardi} {et~al.}(2010){Bernardi}, {de Bruyn}, {Harker},
  {Brentjens}, {Ciardi}, {Jeli{\'c}}, {Koopmans}, {Labropoulos}, {Offringa},
  {Pandey}, {Schaye}, {Thomas}, {Yatawatta}, \& {Zaroubi}}]{bernardi_2010}
{Bernardi}, G., {de Bruyn}, A.~G., {Harker}, G., {et~al.} 2010, \aap, 522, A67

\bibitem[{{Bernardi} {et~al.}(2011){Bernardi}, {Mitchell}, {Ord}, {Greenhill},
  {Pindor}, {Wayth}, \&
  {Wyithe}}]{bernardi_subtraction-of-point-sources-in-radio-astro_2011}
{Bernardi}, G., {Mitchell}, D.~A., {Ord}, S.~M., {et~al.} 2011, \mnras, 413,
  411

\bibitem[{{Brentjens}(2011)}]{brentjens_2011}
{Brentjens}, M.~A. 2011, \aap, 526, A9

\bibitem[{Brentjens \& de~Bruyn(2005)}]{brentjens_faraday_2005}
Brentjens, M.~A. \& de~Bruyn, A.~G. 2005, Astronomy and Astrophysics, 441, 1217

\bibitem[{{Briggs} \& {Cornwell}(1992)}]{briggs_1992}
{Briggs}, D.~S. \& {Cornwell}, T.~J. 1992, in Astronomical Society of the
  Pacific Conference Series, Vol.~25, Astronomical Data Analysis Software and
  Systems I, ed. D.~M. {Worrall}, C.~{Biemesderfer}, \& J.~{Barnes}, 170

\bibitem[{{Clark}(1980)}]{1980_clark}
{Clark}, B.~G. 1980, \aap, 89, 377

\bibitem[{{Cotton} \&
  {Uson}(2008)}]{cotton_pixelization-in-interferometry_2008}
{Cotton}, W.~D. \& {Uson}, J.~M. 2008, \aap, 490, 455

\bibitem[{El-Behery \& MacPhie(1980)}]{el-behery_1980}
El-Behery, I. \& MacPhie, R. 1980, Antennas and Propagation, IEEE Transactions
  on, 28, 234

\bibitem[{{Feain} {et~al.}(2009){Feain}, {Ekers}, {Murphy}, {Gaensler},
  {Macquart}, {Norris}, {Cornwell}, {Johnston-Hollitt}, {Ott}, \&
  {Middelberg}}]{feain_2009}
{Feain}, I.~J., {Ekers}, R.~D., {Murphy}, T., {et~al.} 2009, \apj, 707, 114

\bibitem[{Frick {et~al.}(2010)Frick, Sokoloff, Stepanov, \&
  Beck}]{frick_wavelet-based_2010}
Frick, P., Sokoloff, D., Stepanov, R., \& Beck, R. 2010, Monthly Notices of the
  Royal Astronomical Society, 401, L24

\bibitem[{{Harvey-Smith} {et~al.}(2010){Harvey-Smith}, {Gaensler}, {Kothes},
  {Townsend}, {Heald}, {Ng}, \& {Green}}]{harvey-smith_2010}
{Harvey-Smith}, L., {Gaensler}, B.~M., {Kothes}, R., {et~al.} 2010, \apj, 712,
  1157

\bibitem[{Heald {et~al.}(2009)Heald, Braun, \& Edmonds}]{heald_westerbork_2009}
Heald, G., Braun, R., \& Edmonds, R. 2009, Astronomy and Astrophysics, 503, 409

\bibitem[{{Heald}(2012)}]{heald_2012}
{Heald}, G.~H. 2012, \apjl, 754, L35

\bibitem[{{H{\"o}gbom}(1974)}]{1974_hogbom}
{H{\"o}gbom}, J.~A. 1974, \aaps, 15, 417

\bibitem[{{Iacobelli} {et~al.}(2012){Iacobelli}, {Haverkorn}, \&
  {Katgert}}]{iacobelli_2012}
{Iacobelli}, M., {Haverkorn}, M., \& {Katgert}, P. 2012, ArXiv e-prints

\bibitem[{{Li} {et~al.}(2011){Li}, {Brown}, {Cornwell}, \& {de
  Hoog}}]{li_compressed-sensing-rmsynth_2011}
{Li}, F., {Brown}, S., {Cornwell}, T.~J., \& {de Hoog}, F. 2011, \aap, 531,
  A126

\bibitem[{{Mao} {et~al.}(2010){Mao}, {Gaensler}, {Haverkorn}, {Zweibel},
  {Madsen}, {McClure-Griffiths}, {Shukurov}, \& {Kronberg}}]{mao_2010}
{Mao}, S.~A., {Gaensler}, B.~M., {Haverkorn}, M., {et~al.} 2010, \apj, 714,
  1170

\bibitem[{{Mao} {et~al.}(2012{\natexlab{a}}){Mao}, {McClure-Griffiths},
  {Gaensler}, {Brown}, {van Eck}, {Haverkorn}, {Kronberg}, {Stil}, {Shukurov},
  \& {Taylor}}]{mao_2012b}
{Mao}, S.~A., {McClure-Griffiths}, N.~M., {Gaensler}, B.~M., {et~al.}
  2012{\natexlab{a}}, \apj, 755, 21

\bibitem[{{Mao} {et~al.}(2012{\natexlab{b}}){Mao}, {McClure-Griffiths},
  {Gaensler}, {Haverkorn}, {Beck}, {McConnell}, {Wolleben}, {Stanimirovi{\'c}},
  {Dickey}, \& {Staveley-Smith}}]{mao_2012}
{Mao}, S.~A., {McClure-Griffiths}, N.~M., {Gaensler}, B.~M., {et~al.}
  2012{\natexlab{b}}, \apj, 759, 25

\bibitem[{{O'Sullivan} {et~al.}(2012){O'Sullivan}, {Brown}, {Robishaw},
  {Schnitzeler}, {McClure-Griffiths}, {Feain}, {Taylor}, {Gaensler},
  {Landecker}, {Harvey-Smith}, \& {Carretti}}]{osullivan_2012}
{O'Sullivan}, S.~P., {Brown}, S., {Robishaw}, T., {et~al.} 2012, \mnras, 421,
  3300

\bibitem[{{Perley}(1999)}]{perley_1999}
{Perley}, R.~A. 1999, in Astronomical Society of the Pacific Conference Series,
  Vol. 180, Synthesis Imaging in Radio Astronomy II, ed. G.~B. {Taylor}, C.~L.
  {Carilli}, \& R.~A. {Perley}, 275

\bibitem[{Schwarz(1978)}]{schwarz_1978}
Schwarz, G. 1978, The annals of statistics, 6, 461

\bibitem[{{Van Eck} {et~al.}(2011){Van Eck}, {Brown}, {Stil}, {Rae}, {Mao},
  {Gaensler}, {Shukurov}, {Taylor}, {Haverkorn}, {Kronberg}, \&
  {McClure-Griffiths}}]{vaneck_2011}
{Van Eck}, C.~L., {Brown}, J.~C., {Stil}, J.~M., {et~al.} 2011, \apj, 728, 97

\bibitem[{{van Weeren} {et~al.}(2012){van Weeren}, {R{\"o}ttgering}, {Intema},
  {Rudnick}, {Br{\"u}ggen}, {Hoeft}, \& {Oonk}}]{vanweeren_2012}
{van Weeren}, R.~J., {R{\"o}ttgering}, H.~J.~A., {Intema}, H.~T., {et~al.}
  2012, \aap, 546, A124

\bibitem[{{Yatawatta}(2010)}]{yatawatta_fundamental_2010}
{Yatawatta}, S. 2010, arXiv:1008.1892

\end{thebibliography}

\Online
\onecolumn
\appendix

\section{Iterative maximization approach\label{sec:Maximization-approach}}

We assume that the optimal model point position is shifted by some
small distance, $\delta \phi_{i},$ relative to the grid point location found
using CLEAN, $\overline{\phi_{i}}$ 
\begin{equation}
\phi_{i}=\overline{\phi_{i}}+\delta \phi_{i}.
\end{equation}

The maximum likelihood solution is obtained by minimizing $H$ in Eq.~\ref{eq:log_likelihood} with respect to $\delta \phi_{i}$, which gives
\begin{equation}
\frac{\partial H(d|M)}{\partial\delta \phi_{i}}=\frac{2}{\sigma_{n}^{2}}\sum_{j}\Re\left[i\lambda^2_{j}m_{i} \left\{m_i^{*} - \left(d_{j} - \widetilde{\widetilde{d_{j}^{i}}}\right)^{*}e^{2 i\lambda^2_{j}\phi_{i}} \right\}\right]=0\label{eq:ML_equation}
\end{equation}
where $\widetilde{\widetilde{d_{j}^{i}}}$ is the same as $\widetilde{d_{j}}$, Eq.~\ref{eq:model_in_data-space}, except that the sum is over $k\neq i$. The operator $\Re$ selects the real part of its argument. This expression cannot be solved analytically for $\delta \phi_{i}$, but we can use it to search for the solutions iteratively. To do so we Taylor expand the exponential term in Eq.~\ref{eq:ML_equation} to first order, thereby making $H$ second order in $\delta \phi_{i}$. This gives
\begin{equation}
\frac{\partial H}{\partial\delta \phi_{i}}=\frac{2}{\sigma_{n}^{2}}\sum_{j}\Re\left[\lambda^2_{j}m_{i}i\left\{m_{i}^{*} - \left(d_{j} - \widetilde{\widetilde{d_{j}^{i}}}\right)^{*}e^{2 i\lambda^2_{j}\overline{\phi_{i}}}-2 i\lambda^2_{j}\left(\widetilde{\widetilde{d_{j}^{i}}}-d_{j}\right)^{*}e^{2 i\lambda^2_{j}\overline{\phi_{i}}}\delta \phi_{i}\right\} \right]=0,
\end{equation}
which we can rewrite in the form

\begin{equation}
A=B\delta \phi_{i},\label{eq:delta_x_solutions}
\end{equation}
where
\begin{align}
A = & \sum_{j}\Re\left[\left\{m_{i}^{*} - \left(d_{j}-\widetilde{\widetilde{d_{j}^{i}}}\right)^{*}e^{2 i\lambda^2_{j}\overline{\phi_{i}}}\right\} \lambda^2_{j}m_{i}i\right],\nonumber \\
B = & \sum_{j}\Re\left[-2 \lambda^4_{j}m_{i}\left(d_{j}-\widetilde{\widetilde{d_{j}^{i}}}\right)^{*}e^{2 i\lambda^2_{j}\overline{\phi_{i}}}\right].
\end{align}

We must also solve for an updated flux. We can again extremize Eq.~\ref{eq:log_likelihood}, this time with respect to $m_{i}$. We find
\begin{equation}
\frac{\partial H}{\partial m_{i}}=\frac{1}{2\sigma_{n}^{2}}\sum_{j}[m_{i}^{*}-(d_{j}-\widetilde{\widetilde{d_{j}}})^{*}e^{2 i \phi_{i}\lambda^2_{j}}]
\end{equation}
and thus

\begin{equation}
m_{i}=\frac{1}{N_d}\sum_{j=1}^{N_d}[(d_{j}-\widetilde{\widetilde{d_{j}^{i}}})e^{-2 i\phi_{i}\lambda^2_{j}}].\label{eq:delta_m_solutions}
\end{equation}
In our example implementation, we solve for $\delta \phi$ and $m$
iteratively until convergence is achieved. We also attempt to merge nearby model components to reduce the degrees of freedom in the model according to the prescription described in Sec.~\ref{sec:The-algorithm}. 

We tried other iterative schemes, e.g. solving for position and flux changes by inverting the Hessian matrix of $H$, but found that the approach given here is the most stable.

\end{document}